\documentclass[twocolumn,10pt,a4paper,byrevtex,floatfix,prl,unsortedaddress,superscriptaddress]{revtex4}

\usepackage{amssymb}
\usepackage{soul}
\usepackage{graphicx}
\usepackage{color}
\usepackage[colorlinks,allcolors=blue]{hyperref}
\usepackage{subfigure}
\usepackage{amsmath}
\usepackage{newtxtext}
\usepackage{newtxmath}
\usepackage{microtype}
\usepackage{bm}
\usepackage[per-mode=symbol]{siunitx}
\usepackage[capitalize]{cleveref}

\definecolor{Red}{rgb}{0,0,0}
\definecolor{DarkGreen}{rgb}{0,0.8,0}

\definecolor{Purple}{rgb}{0.7,0,1}
\newcommand{\cesar}[1]{\textcolor{Red}{#1}}
\definecolor{Orange}{rgb}{1,0.6,0}
\newcommand{\diego}[1]{\textcolor{Red}{#1}}

\definecolor{Blue}{rgb}{0,0,1}
\newcommand{\ali}[1]{\textcolor{Red}{#1}}
\definecolor{Pink}{rgb}{1,0,1}

\begin{document}

\title{Observation of magnetic state dependent thermoelectricity in superconducting spin valves}

\author{C\'esar Gonz\'alez-Ruano$^{\dag}$}
\affiliation{Departamento F\'isica de la Materia Condensada C-III, Universidad Aut\'onoma de Madrid, Madrid 28049, Spain}

\author{Diego Caso$^{\dag}$}
\affiliation{Departamento F\'isica de la Materia Condensada C-III, Universidad Aut\'onoma de Madrid, Madrid 28049, Spain}

\author{Jabir Ali Ouassou}
\affiliation{Center for Quantum Spintronics, Department of Physics, Norwegian University of Science and Technology, NO-M7T9Q Trondheim, Norway}

\author{Coriolan Tiusan}
\affiliation{Department of Solid State Physics and Advanced Technologies, Faculty of Physics, Babes-Bolyai University, Cluj Napoca 400114, Romania}
\affiliation{Institut Jean Lamour, Nancy Universit\`{e}, 54506 Vandoeuvre-les-Nancy Cedex, France}

\author{Yuan Lu}
\affiliation{Institut Jean Lamour, Nancy Universit\`{e}, 54506 Vandoeuvre-les-Nancy Cedex, France}

\author{Jacob Linder}
\affiliation{Center for Quantum Spintronics, Department of Physics, Norwegian University of Science and Technology, NO-M7T9Q Trondheim, Norway}

\author{Farkhad G. Aliev}
\email[e-mail: ]{farkhad.aliev@uam.es}
\affiliation{Departamento F\'isica de la Materia Condensada C-III, Instituto Nicol\'as Cabrera (INC) and  Condensed Matter Physics Institute (IFIMAC), Universidad Aut\'onoma de Madrid, Madrid 28049, Spain}

\begin{abstract}
Superconductor-ferromagnet tunnel junctions demonstrate giant thermoelectric effects which are being exploited to engineer ultra-sensitive terahertz radiation detectors. Here, we experimentally observe the recently predicted complete magnetic control over thermoelectric effects in a superconducting spin valve, including the dependence of its sign on the magnetic state of the spin valve. The description of the experimental results is improved by the introduction of an interfacial domain wall in the spin filter layer interfacing the superconductor. Surprisingly, the application of high in-plane magnetic fields induces a double sign inversion of the thermoelectric effect, which exhibits large values even at applied fields twice the superconducting critical field. 
\end{abstract}

\maketitle

\section{Introduction}

The competition between superconductivity (S) and ferromagnetism (F) can under certain conditions result in a synergy of these otherwise antagonistic states \cite{Buzdin2005}.
In recent years, a variety of exotic phenomena have been demonstrated in devices that exploit this synergy.
Notable examples include long-ranged spin-triplet supercurrents \cite{keizer_nature_06,robinson_science_10,khaire_prl_10}, spin-valve Josephson junctions \cite{Birge2018}, superconducting spin-valves with record-high magnetoresistance, and the giant thermoelectric (TE) effect.
These effects are considered as potential ingredients in the next generation of low-dissipation cryogenic devices \cite{Yang2021, Madden2018,Ai2021}.

In general, there exists considerable interest in identifying material platforms for improved TE devices. At low temperatures, TE effects are expected to be vanishingly small in both normal metals and bulk superconductors. Instead, they have been investigated mainly in superconductor/normal-metal hybrids, where they have been used in micro-refrigeration and thermometry \cite{Giazotto2009}. More recently, fascinating theoretical predictions \cite{Machon2013,Machon2014,Ozaeta2014, Kalenkov2014,Dutta2017,Savander2020} have opened the door to unexplored spin-dependent TE effects in S/F hybrids. The transport of spin and charge due to temperature gradients in such systems have only been investigated experimentally in a few works \cite{Kolenda2016,Kolenda2016_2,Kolenda2017,Heidrich2019}. Kolenda \textit{et al.}~\cite{Kolenda2016} reported on the experimental observation of an enhanced Seebeck coefficient (up to \SI{100}{\micro\volt\per\kelvin}) when a large magnetic field of \SI{1}{\tesla} splits the quasiparticle band structure of a superconductor.
When combined with a spin-filtering interface, this spin splitting breaks the electron-hole symmetry, producing the observed ``giant TE effect'', which is now being exploited to develop ultrasensitive radiation detectors \cite{Heikkila2018}. 

So far, the experimental tuning of giant TE effects in S/F hybrids has been performed either by applying large magnetic fields \cite{Kolenda2016} or by exchange coupling a superconductor to a ferromagnetic insulator \cite{Heikkila2018}. Recently, however, a different method to control the TE effect 
has been predicted in superconductor/ferromagnet/ferromagnetic insulator systems \cite{Ouassou2022}.
This method can turn the superconducting TE effect on and off \textit{in situ}, as well as reversing its sign. Here, by interfacing a superconductor with a spin valve with large spin filtering capability, we experimentally demonstrate the mentioned complete magnetic control of the superconducting TE effect. 
This includes evidence of an antisymmetric TE effect, where a change of the magnetic state of the spin valve inverts the direction of the TE current. 
Controlling the sign of the thermopower, analogous to the inversion of TE signals between $p$- and $n$-doped semiconductors, enables the design of Peltier elements based on superconducting spin valves.

\begin{figure}
\begin{center}
\includegraphics[width=\linewidth]{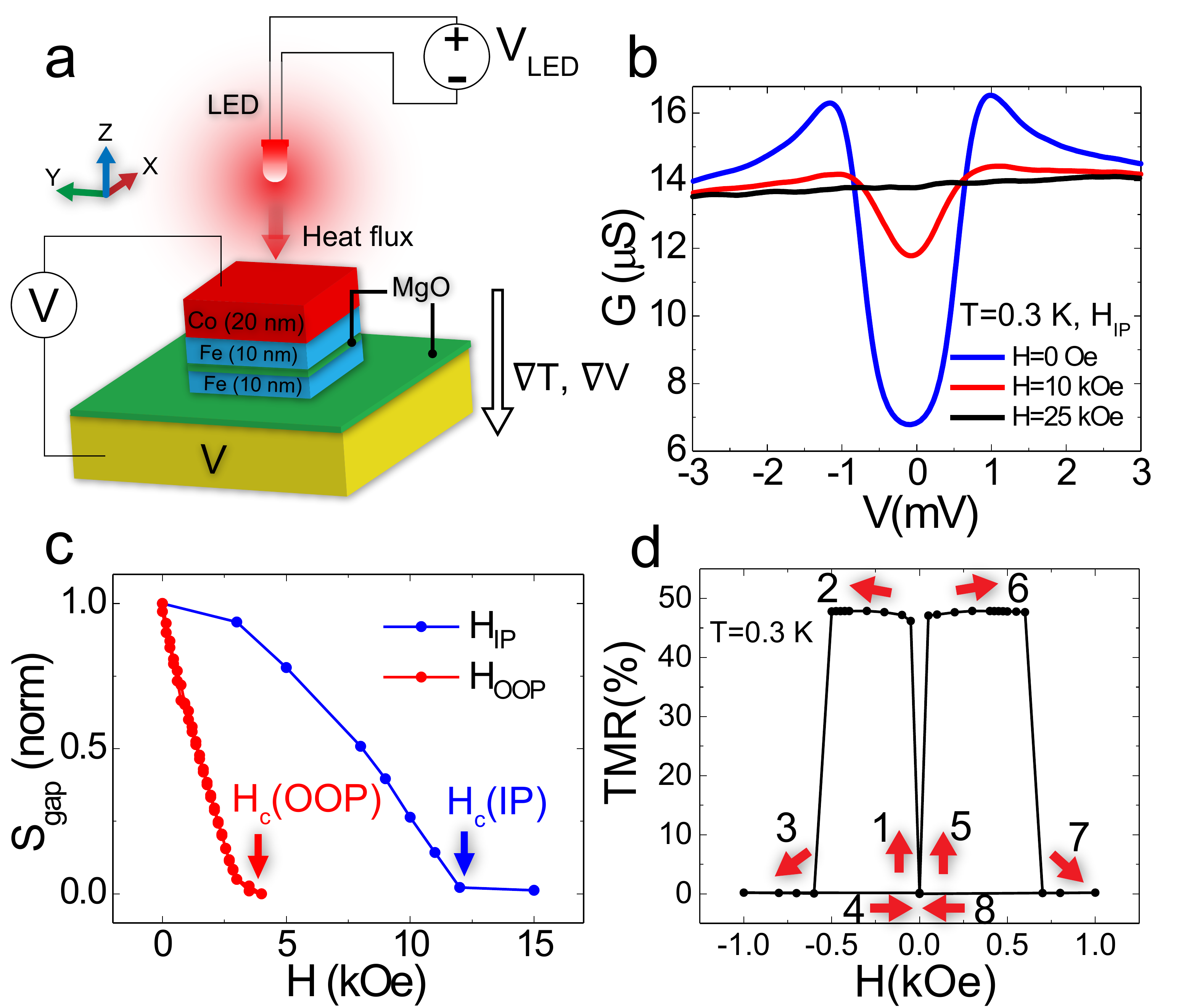}
\caption{(a)~Sketch of the S/F/F junctions when heated by a LED. (b)~Typical conductance--bias curves measured at $0.07T_c$ at three different applied IP magnetic fields. (c)~Normalized superconducting gap depth (S$_\text{gap}$) taken from $G(V)$ curves vs applied IP and OOP magnetic fields. (d)~Typical tunneling magnetoresistance curves measured with IP magnetic field at $V=5$~mV and $T=0.3$~K. The numbers indicate the order of the field sweeping and resistance changes.}
\label{Fig1}
\end{center}
\end{figure}

\section{Experimental results}

\Cref{Fig1}(a) illustrates the main
experimental set-up and junctions investigated. 
We have measured TE effects in V(40)/MgO(2)/Fe(10)/MgO(2)/Fe(10)/Co(20) single-crystalline junctions epitaxially grown on MgO(001) substrates, with the thickness of each layer given in nanometers in parentheses. More details about the junctions, experimental set-up and procedures can be found in the Supplemental Material\ref{supplemental}.
Here, V is a BCS superconductor, Fe and Co are ferromagnetic metals, and MgO is a symmetry filtering insulator.
The magnetically hard Fe/Co electrode allows a precise detection of the orientation of the magnetically free Fe layer interfacing the superconductor. \Cref{Fig1}(b)-(c) presents a general electron transport characterization of the junctions in the superconducting state as a function of the applied bias and external magnetic field. The superconducting gap in the V electrode is suppressed by in-plane (IP) and out-of-plane (OOP) fields of about \SI{1.7}{\tesla} and \SI{0.4}{\tesla}, respectively (Fig. \cref{Fig1}(c)).
The magnetoresistance values of 35-55\% provide an estimation of the effective spin polarization of the Fe electrodes of around 0.75--0.85 for the different junctions studied based on the Slonczewski model, which was adapted for the case of two resistances in series, one of which depends on the relative magnetization angle and spin polarization. This procedure and results are line with previous reports \cite{Martinez2018,GonzalezRuano2020}.
At first glance, these TMR values of $\sim45\%$ may seem very low compared to crystalline F/F junctions where values from $180\%$ to $300\%$ have been reported \cite{Parkin2004,Yuasa2004}. However, this value is not related to the quality of the crystalline structure (note for example the polarization values in figure \ref{Fig4}b, exceeding 0.8), but rather with the fact that the structure of these junctions is N/F/F, including a normal metal electrode (vanadium) with a second tunnel barrier, which has an almost fixed resistance and strongly hampers the total TMR ratio. Indeed, in experiments in control F/F junctions grown in the same conditions, we have previously observed TMR values from $185\%$ to $330\%$ \cite{Guerrero2007,Gonzalez-Ruano2021}. Further evidence for the high crystalline quality and effective spin filtering in our V/MgO/Fe/MgO/Fe/Co junctions comes from their record high tunneling magnetoresistance achieved at biases exceeding 1~V \cite{Gonzalez-Ruano2021}.

To study the TE effect as a function of the spin-valve state, the soft Fe layer was rotated while the hard Fe/Co layer remained fixed.
We did this by applying a rotating in-plane magnetic field, with a magnitude between the coercive fields of the two magnetic layers.
For the hard layer the coercive field is typically larger than 500~Oe (\cref{Fig1}(d)), while for the soft layer it's smaller than 50 Oe, determined by the magneto-crystalline anisotropy.
This procedure guarantees a reorientation between the parallel (P), antiparallel (AP) and perpendicular in-plane (PIP) configurations of the spin valve. 
For each $3^\circ$ rotation of the applied field, the temperature gradient~${\nabla}T$ was re-established via the LED heater, and the resulting TE response ${\Delta}V$ was measured.

\begin{figure}
\begin{center}
\includegraphics[width=\linewidth]{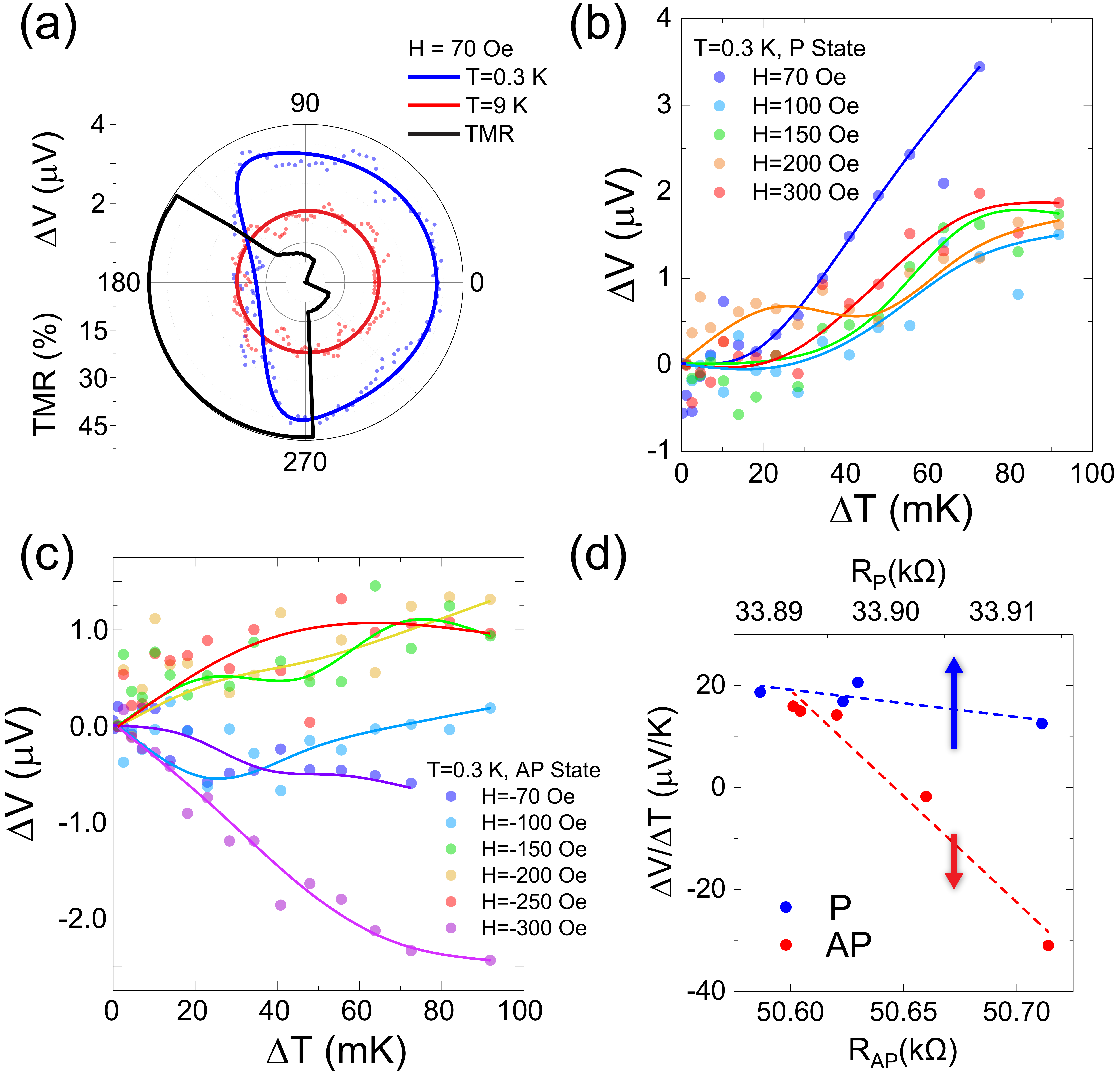}
\caption{Thermoelectric response of a S/F/F junction measured under a rotation and fixed values of the applied magnetic field at $T=0.3$~K. (a) TE response at $H=70$~Oe for in-plane rotations of the magnetic field at $V_\text{LED}=7.3$~V (${\Delta}T\approx113$~mK), below and above $T_c$. The tunnel magnetoresistance of the spin-valve is also displayed against the rotation angle. (b) Response in the P configuration of the spin-valve stack. (c) Same experiment for the AP configuration. The TE voltage changes its sign and intensity depending not directly on the applied field, but on the saturation of the soft FM layer. The related analysis is shown in panel (d), where the average value of the TE voltage is plotted against the measured resistance for the P (blue, upper horizontal axis) and AP (red, lower horizontal axis) configurations. For the P state, a lower resistance implies a better polarization, while in the AP state the polarization is better with a higher resistance.}
\label{Fig2}
\end{center}
\end{figure}

Fig. \ref{Fig2}(a) shows the variation of the TE voltage generated during a magnetization rotation of the free layer under an in-plane applied magnetic field of 70~Oe. While the transition between the P and PIP states hardly affects the values of the TE response, a strong reduction of the TE voltage of more than a factor of 2 is observed when the free layer becomes close to AP to the fixed Fe/Co layer. Note that there is a slight latency of about 10--15$^\circ$ between the angle of the applied magnetic field and the real average magnetization orientation of the Fe layer. This is a natural consequence of the experimental process: the Fe magnetization has to overcome the magneto-crystalline anisotropy to follow the slowly rotating applied field (see for example Ref. \cite{Gabor2011}). The complete $360^\circ$ rotation takes about two hours, since we stop at each intermediate angle to measure the TE response. Figures \cref{Fig2}(b) and c show the TE response of the samples vs the induced temperature gradient for different magnetic configurations and applied fields. Note that these temperature gradients have been estimated by first simulating the response to the incoming heat flux and subsequently recalculating ${\Delta}T$ from $V_\text{LED}$ based on the LED calibration curves (see Supplemental Material\ref{supplemental}).
In the P state, changing the applied field by less than a few hundred Oe does not qualitatively change the TE response (\cref{Fig2}(b)). However, in the AP state, varying the magnetic field has a dramatic effect on the TE response and can even change its sign as seen in \cref{Fig2}(c). We note that no asymmetry of ${\Delta}V$ or dependence on $V_\text{LED}$ was observed above $T_c$ (\cref{Fig2}(a)). Control experiments on short-circuited junctions also revealed at least an order of magnitude drop of the TE response regardless of the magnetic state (see Supplemental Material\ref{supplemental}).

In order to understand the possible reasons for the TE sign change in the AP state, we analyzed (\cref{Fig2}(d)) the TE response obtained for a fixed temperature gradient as a function of the resistance in the P and AP states obtained during each particular TE experiment at different applied in-plane magnetic fields (not exceeding 500~Oe). While the TE response in the P state is rather robust to the variation of the resistance (i.e. presence of magnetic inhomogeneities), in the AP state it changes sign with the \emph{reduction} of the influence of these magnetic textures (an increase of the resistance means better magnetization saturation).

\begin{figure}
\begin{center}
\includegraphics[width=\linewidth]{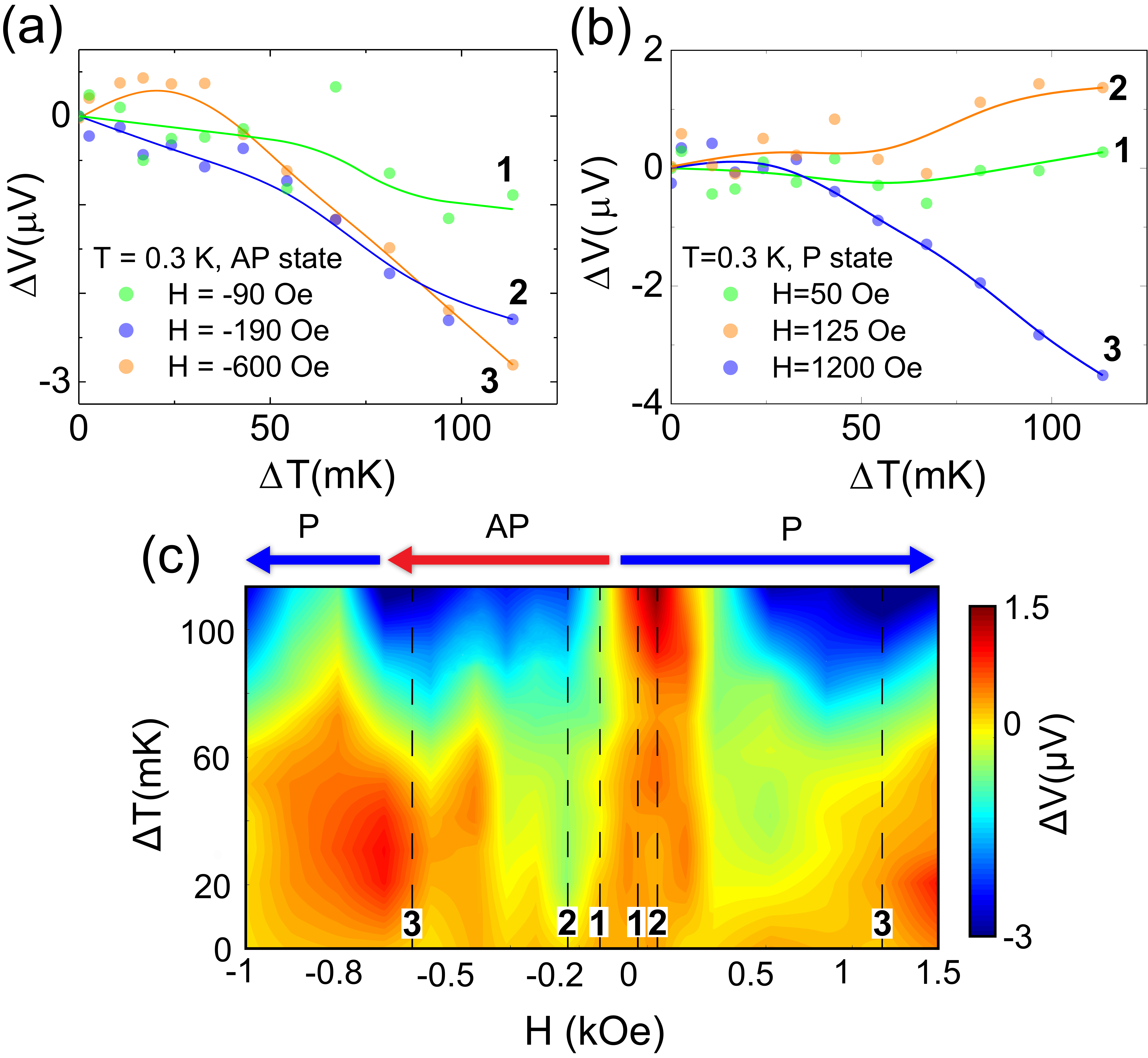}
\caption{Thermoelectric voltage of an S/F/F junction demonstrating sign inversion at high fields in the \diego{AP} state \textbf{(a)} and \diego{P} state \textbf{(b)}. (c)~Colormap of the recorded TE voltage ${\Delta}V$ as a function of the temperature difference ${\Delta}T$ and the applied in-plane field $H$ for the same junction, indicating the P and AP states.}
\label{Fig3}
\end{center}
\end{figure}

\begin{figure}
\begin{center}
\includegraphics[width=\linewidth]{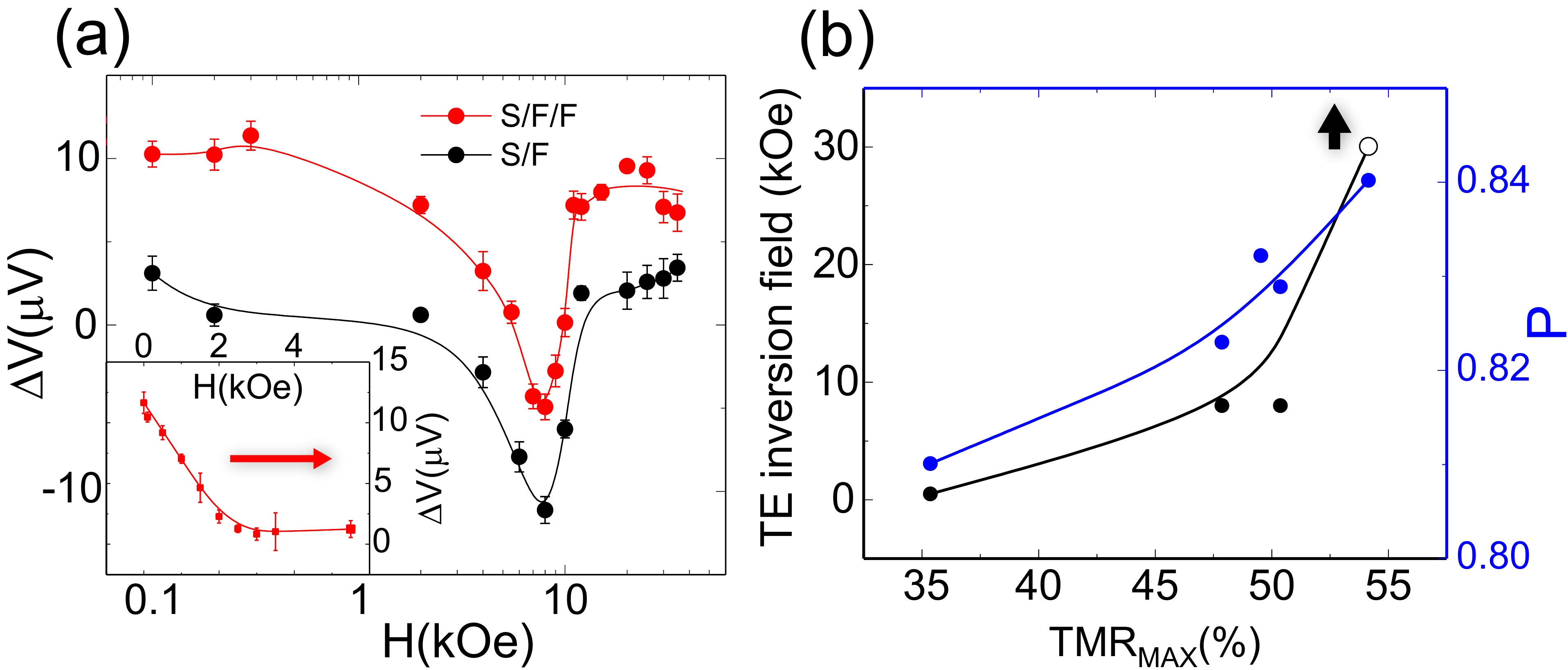}
\caption{(a) Thermoelectric voltage of an S/F/F and S/F junction at $V_\text{LED}=7.3$~V (${\Delta}T\approx113$~mK) vs. applied in-plane field at $T=0.3$~K. The high-field sign inversion is achieved in both samples, and the TE effect is maintained with increasing in-plane fields up to 30~kOe. The inset displays the measured TE voltage in the S/F/F junction under out-of-plane field. At $H=H_c$, superconductivity and its associated TE voltage vanish. (b) TE inversion field and polarization against the maximum TMR value for all the S/F/F junctions under study.}
\label{Fig4}
\end{center}
\end{figure}

Interestingly, some junctions revealed a TE sign inversion both in the AP state and also under a sufficiently high applied in-plane magnetic field in the P state (\cref{Fig3}). The TE response in the P state is positive and robust at fields below 0.5~kOe (\cref{Fig3}(a)), and becomes negative for higher magnetic fields. In contrast, in the AP state the TE response is already negative for much smaller fields (\cref{Fig3}(b)), right after the spin-valve switched into the AP state (see steps 1-2 in \cref{Fig1}(d)). Further increasing the negative magnetic field (path 2-3 in \cref{Fig1}(d)) reorients the hard layer so the spin-valve is again in the P configuration, and again the sign of the TE effect follows the same trend as for positive fields. \Cref{Fig3}(c) summarizes these observations with a 3D color plot of the TE voltage signal against the applied field and evaluated temperature gradient.
Fig.~\ref{Fig3}(a) also shows that $\Delta V$ could change sign as a function of $\Delta T$ at fixed magnetic field~$H$. One potential explanation could be a temperature-induced change in the interfacial domain wall structure, as discussed in the Supplemental Material\ref{supplemental}.

We have found that the TE voltage sign inversion in the P state under high in-plane magnetic fields is a rather robust effect and is followed by a second TE sign inversion towards positive values when the magnetic field is further increased (\cref{Fig4}(a)). Surprisingly, even for maximum applied in-plane magnetic fields, twice exceeding the second critical magnetic field (compare \cref{Fig1}(c) and \cref{Fig4}(a)), the TE voltage remains high and without clear signatures of diminishing.  A qualitatively similar response has been observed in single barrier V/MgO/Fe junctions, i.e. without the sensing Fe/Co layer. Fig. \ref{Fig4}(b) compares the TE signal inversion field with the TMR and effective spin polarization values of each corresponding sample. Apparently, a higher TMR and correspondingly polarization values shift the TE inversion field range outside our experimental capabilities (35~kOe). This suggests a possible link between the observed effect and interfacial domain wall forming in the Fe electrode. Further experiments are needed to understand the physical mechanism behind the high field TE effects.

\section{Theoretical modelling}

To better understand the physics behind the experimental observations, we explored the setup in \cref{Fig1}(a) via numerical simulations. We employed the Usadel formalism \cite{Bergeret.2018,Chandrasekhar.2008,Belzig.1999,Rammer.1986,Usadel.1970} which describes superconductivity in diffusive heterostructures, together with spin-dependent tunneling boundary conditions \cite{Ouassou.2017,Eschrig.2015lmf,Machon2013,Cottet.2009,Cottet.2007} valid for arbitrary spin polarizations. To numerically solve these equations, we used the Ricatti parametrization \cite{Jacobsen.2015,Schopohl.1998} to calculate spectral properties and a distribution-trace parametrization \cite{Ouassou.2018fbw,Bergeret.2018,Chandrasekhar.2008,Belzig.1999} to calculate the nonequilibrium transport properties. The theoretical and numerical approach is described in more detail in Ref.~\cite{Ouassou2022}.

\begin{figure}[b!]
\begin{center}
\includegraphics[width=\columnwidth]{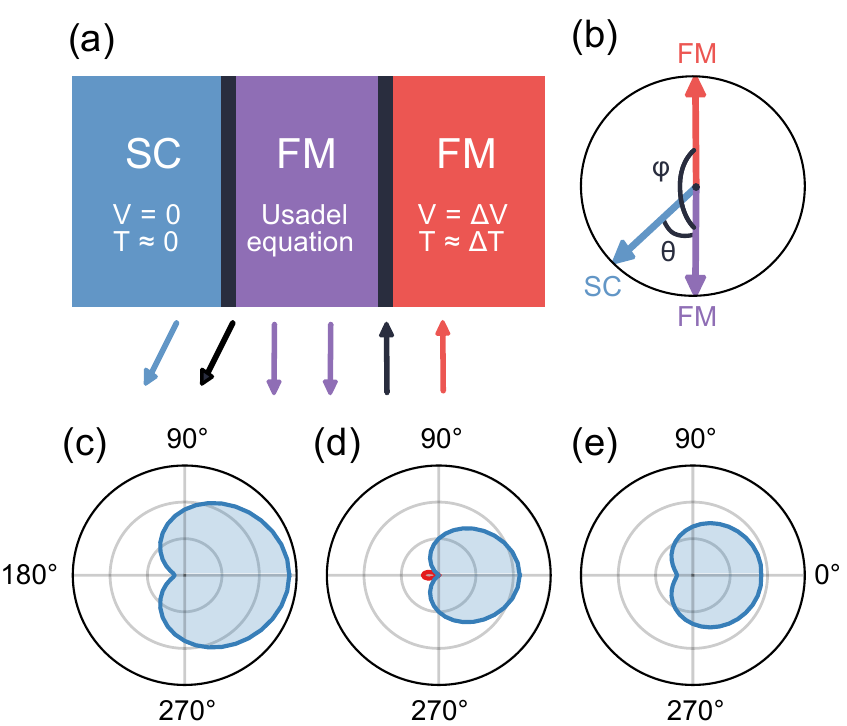}
\caption{(a)~Numerical model, including field directions (shown here for the AP configuration). (b)~Definitions of angles $\theta$ and $\varphi$ with respect to the field directions in the model. (c--e) Numerical results for the magnetically dependent thermoelectric voltage ${\Delta}V(\varphi)$. Blue and red correspond to positive and negative voltages, while the radius in each plot is $|{\Delta}V|=0.04\Delta/e$ where $e$ is the elementary charge. The three plots correspond to (c)~$\theta=0, P_1=P_2=80\%;$ (d)~$\theta=0, P_1=60\%, P_2=80\%;$ (e)~$\theta=\pi/4, P_1=60\%, P_2=80\%.$}
\label{fig:model}
\end{center}
\end{figure}

The numerical model used herein is sketched in \cref{fig:model}(a). The superconductor~(V) was treated as a BCS superconducting reservoir with an effective spin-splitting $h=\Delta/10$, near-zero temperature $T=T_c/100$, and electrical grounding $V=0$. The hard ferromagnet (Fe/Co) was treated as a non-superconducting metallic reservoir at an elevated temperature $T=T_c/2$ and voltage $V={\Delta}V$. The interfaces to the soft ferromagnet (V/MgO/Fe and Fe/MgO/Fe/Co) were treated using spin-polarized tunneling boundary conditions with spin polarizations $P_1,P_2$ and a low tunneling conductance $G_T=G_D/5$ where $G_D$ is the Drude conductance of the soft ferromagnet. These parameters model the high spin filtering capabilities and low transparencies of the MgO barriers. In the soft ferromagnet (Fe), we used an exchange splitting $h=30\Delta$. For each magnetic configuration, we (i)~solved the Usadel equation for 80 different $e{\Delta}V/\Delta\in[-0.04,+0.04]$; (ii)~used this to calculate the current $I({\Delta}V)$; (iii)~interpolated the open-circuit voltage from $I({\Delta}V)\equiv0$. This yielded the TE voltage as function of magnetic configuration.

The magnetic configurations we considered are illustrated in \cref{fig:model}(a--b). We take the hard ferromagnet (red) to be oriented along one in-plane axis (up), and the spin filtering at the MgO barrier is assumed to be parallel to this orientation (black). The soft ferromagnet is then rotated by an in-plane angle~$\varphi$ relative to the hard ferromagnet (purple), which here is sketched for the antiparallel case $\varphi=\pi$. At the superconductor interface, we include the possibility for an interfacial domain wall described by an out-of-plane angle~$\theta$. This affects both the spin filtering at the second MgO barrier (black) and the direction of spin splitting inside the neighboring superconductor (blue). \cref{fig:model}(c--e) show the numerical results for the TE voltage~${\Delta}V(\varphi)$ across the junction as function of the in-plane misalignment angle between the two ferromagnets. In panel~(c), we see the predicted response for a junction with two identical spin filters ($P_1=P_2$) and no interfacial domain wall ($\theta=0$). We then predict an \emph{asymmetric TE effect}, by which we mean that ${\Delta}V(0)$ is maximal while ${\Delta}V(\pi)\rightarrow0$. This can intuitively be understood as follows. If the two spin filters are identical ($P_1=P_2$ and $\varphi=0$), then ``filtering the spins twice'' does not significantly change the physics compared to having only one spin filter. The latter case has in previous work been shown to produce a giant TE effect in superconductor/ferromagnet systems \cite{Machon2013,Machon2014,Ozaeta2014,Kalenkov2014,Dutta2017,Savander2020,Kolenda2016,Kolenda2016_2,Kolenda2017,Heidrich2019}. On the other hand, the effects of two identical but oppositely aligned spin filters cancel, so any TE effect due to spin filtering should vanish for $\varphi=\pi$. In panel~(d), we see the case of different spin filters ($P_1<P_2$). It is now possible for one ferromagnet to dominate the spin splitting of the superconducting density of states, while the other dominates the spin filtering process. Via the mechanism explored in detail in \cite{Ouassou2022}, this leads to an antisymmetric contribution to ${\Delta}V(\varphi)$, whereby ${\Delta}V(0)$ and ${\Delta}V(\pi)$ have opposite signs. In the extreme case of $P_1~{\ll}~P_2$ the result is a purely antisymmetric shape for ${\Delta}V(\varphi)$, whereas for $P_1,~P_2$ of similar magnitude the theory predicts $|{\Delta}V(\pi)|\ll|{\Delta}V(0)|$. In panel~(e), we show the effect of adding an out-of-plane interfacial domain wall to panel~(d), which clearly suppresses the antisymmetric contribution.

While the simplest model presented in \cref{fig:model}(c) captures the essential features of \cref{Fig2}(a), including an interfacial domain wall in the model enhances the agreement with the experiment. Specifically, in the absence of externally applied fields, such a domain wall would produce the results in panel~\cref{fig:model}(e), which also agrees well with \cref{Fig2}(a). However, as the in-plane applied field is ramped up, the domain wall should be rotated into the thin-film plane: $\theta\rightarrow0$. In this case, we would gradually move towards panel~(d), where ${\Delta}V(\varphi)$ changes sign. This qualitatively agrees with the experimental observations in \cref{Fig2}d, where it is found that ${\Delta}V(\pi)$ changes sign for increasing magnetic saturation while ${\Delta}V(0)$ changes only slightly. For plots of $\Delta V$ as a function of $\theta$, thus modeling the magnetic field dependence of the thermoelectric effect, see the Supplemental Material\ref{supplemental}.

\section{Discussion and Conclusions}

While our numerical modelling qualitatively explains the experiments at low magnetic fields where the switching between the P and AP states takes place (\cref{Fig2,Fig3,fig:model}), it does not account for the unexpected strong variation of the TE response in the high field limit, where a double sign change takes place regardless of the presence of the magnetically hard layer (\cref{Fig4}). This is because the spin-resolved particle-hole asymmetry in quasiclassical theory is only present in the superconducting state. A possible factor which may influence the high field TE response is a transformation of the interfacial magnetism at the V/MgO interface \cite{Rau1986} under an applied magnetic field. Initially predicted by numerical simulations \cite{Grempel1980}, spin fluctuations and/or surface atomic layer magnetism in V have been under debate for decades now \cite{Walker1994,Fuchs1996,Bryk2000,Lacina2007,Rubio-Ponce2020}. In our experiments, a sufficiently large in-plane magnetic field could transform the V/MgO interface into an additional, atomically-thin magnetic layer. The induced surface magnetism might strongly affect the exchange splitting of the electron bands in V. The explanation of the high-field TE response behavior remains an intriguing open problem.

\textit{In conclusion}, we report on the experimental control of the superconducting thermoelectric effect using a spin-valve device with a spin filter. We demonstrate both experimentally and by numerical simulations the transition from a strongly asymmetric to an anti-symmetric response depending on the saturation of the AP alignment of the spin-valve, which is likely modulated by an interfacial domain wall. Furthermore, our results point towards an unexpected thermoelectric response in superconductor/ferromagnet junctions under high in-plane magnetic fields. More detailed experimental and theoretical studies are required to understand this behaviour.

\section*{Acknowledgements}
Authors thank Michel Hehn for discussions and help with samples preparation. The work in Madrid was supported by Spanish Ministry of Science and Innovation (PID2021-124585NB-C32 and TED2021-130196B-C22) and Consejer\'ia de Educaci\'on e Investigaci\'on de la Comunidad de Madrid (NANOMAGCOST-CM Ref. P2018/NMT-4321) Grants. F.G.A. also acknowledges financial support from the Spanish Ministry of Science and Innovation through the Mar\'ia de Maeztu Programme for Units of Excellence in R\&D (CEX2018-000805-M) and ``Acci\'on financiada por la Comunidad de Madrid en el marco del convenio plurianual con la Universidad Aut\'onoma de Madrid en L\'inea 3: Excelencia para el Profesorado Universitario''. The work in Trondheim was supported by the Research Council of Norway through grant 323766, and its Centres of Excellence funding scheme grant 262633 ``QuSpin''. J. L. and J. A. O. also acknowledge resources provided by 
Sigma2---the National Infrastructure for High Performance Computing and Data Storage in Norway. C. T. acknowledges  the UEFISCDI project ``MODESKY''ID PN-III-P4-ID- 880 PCE-2020-0230-P, grant No. UEFISCDI: PCE 245/02.11.2021.
\bigskip

$^{\dag}$C.G.-R and D.C. contributed equally to the
manuscript.

\section{Supplemental material}\label{supplemental}

\section*{1. Samples and experimental set-up}

The superconductor--spin-valve multilayer stacks have been grown by molecular beam epitaxy (MBE) in a chamber with a base pressure of $5\times10^{-11}$~mbar while the crystalline quality was controlled by in-situ RHEED measurements, following the procedure described in Ref. \cite{Tiusan2007}. The resulting layered structures were then lithographed into squared samples with lateral sizes ranging from $30\times30$ to $60\times60~\mu\text{m}^2$.

The measurements are performed inside a JANIS$^{\tiny{\textregistered}}$ He$^3$ cryostat (minimum attainable temperature is 0.3~K). The magnetic field is varied using a 3D vector magnet consisting of one solenoid (X axis) with $H_\text{max}=3.5$~T and two Helmholtz coils (Y and Z axis) with $H_\text{max}=1$~T. 

\section*{2. Experimental procedures for the TE response measurements}

In order to produce a temperature gradient in the samples, a commercial Light Emitting Diode (LED) was placed above the sample (see fig. \ref{Fig1} of the main text). Voltage was supplied to the LED (model LUXEON 3030 2D) by a Keithley 228A voltage source. The LED started to emit light at an applied bias of 5.6~V at room temperature. As no direct visual contact could be established inside the cryostat, the thermometer and voltage source were used to check the LED functioning at low temperatures: First, the thermometer closer to the samples showed a steady increase in temperature when the applied voltage was above 6~V. Second, the compliance indicator present in the Keithley voltage source showed the same signal at this voltage than it did at room temperature when the LED was on.

Once this was established, the TE voltage (${\Delta}V$) is measured as follows: first, the resistance and temperature of the sample are measured, and then voltage measurements are taken with an applied current of $I=0$~nA (using the smallest available current range in the Keithley 220 current source, which has a maximum compliance current of $I=1.9995$~nA and a minimum step of 500~fA with an accuracy of $\pm2$~pA). Then, the LED is turned on with the desired voltage ($V_\text{LED}$), and a second voltage measurement is taken at zero current after 2 seconds, before the LED is turned off again. We checked that the voltage readings were stable for waiting times between 1 and 10 seconds after turning on the LED. We chose the 2 seconds because it ensures the gradient temperature without overheating the sample during the experiments. Each voltage measurement is taken using the smallest measuring range of the DMM-552 voltmeter card, and averaging over 100 voltage readings.

We have carried out three types of thermoelectric measurement experiments. First, the magnetic field rotations that have been explained in the experimental results section. Secondly, the TE response of the spin-valve (S/F/F) samples was studied separately in each magnetic configuration. This was done by setting the desired magnetic orientation (i.e. parallel (P) or antiparallel (AP) state) at a fixed temperature (usually below $T_c$ at $T=0.3$~K or above $T_c$ at 9~K) and field magnitude, and measuring the TE voltage ${\Delta}V$ for increasing values of the LED heater voltage $V_\text{LED}$. The sample was let to cool down for a minute after each heating experimental point in order to avoid overheating. The whole process was repeated 5 to 10 times and averaged for each $V_\text{LED}$ to reduce the uncertainty. This was made for different magnetic orientations of the sample, as shown in figs. \ref{Fig2} and \ref{Fig3} on the main text, where instead of $V_\text{LED}$, the corresponding estimated temperature difference ${\Delta}T$ between the Fe and V electrodes is shown (see details on the temperature gradient evaluation below). 

The third type of experiments study the TE response as a function of the applied magnetic field direction for fixed values of $V_\text{LED}$. In all of the experiments, the low bias resistance (at about 5~mV) was also measured, allowing for a precise detection of the magnetization configuration of the free Fe layer. 
More details on the electric transport measurements, along with the samples characterization, can be found in Ref. \cite{Gonzalez-Ruano2021OOP}

\section*{3. Background voltage removal}

\begin{figure}
\begin{center}
\includegraphics[width=\linewidth]{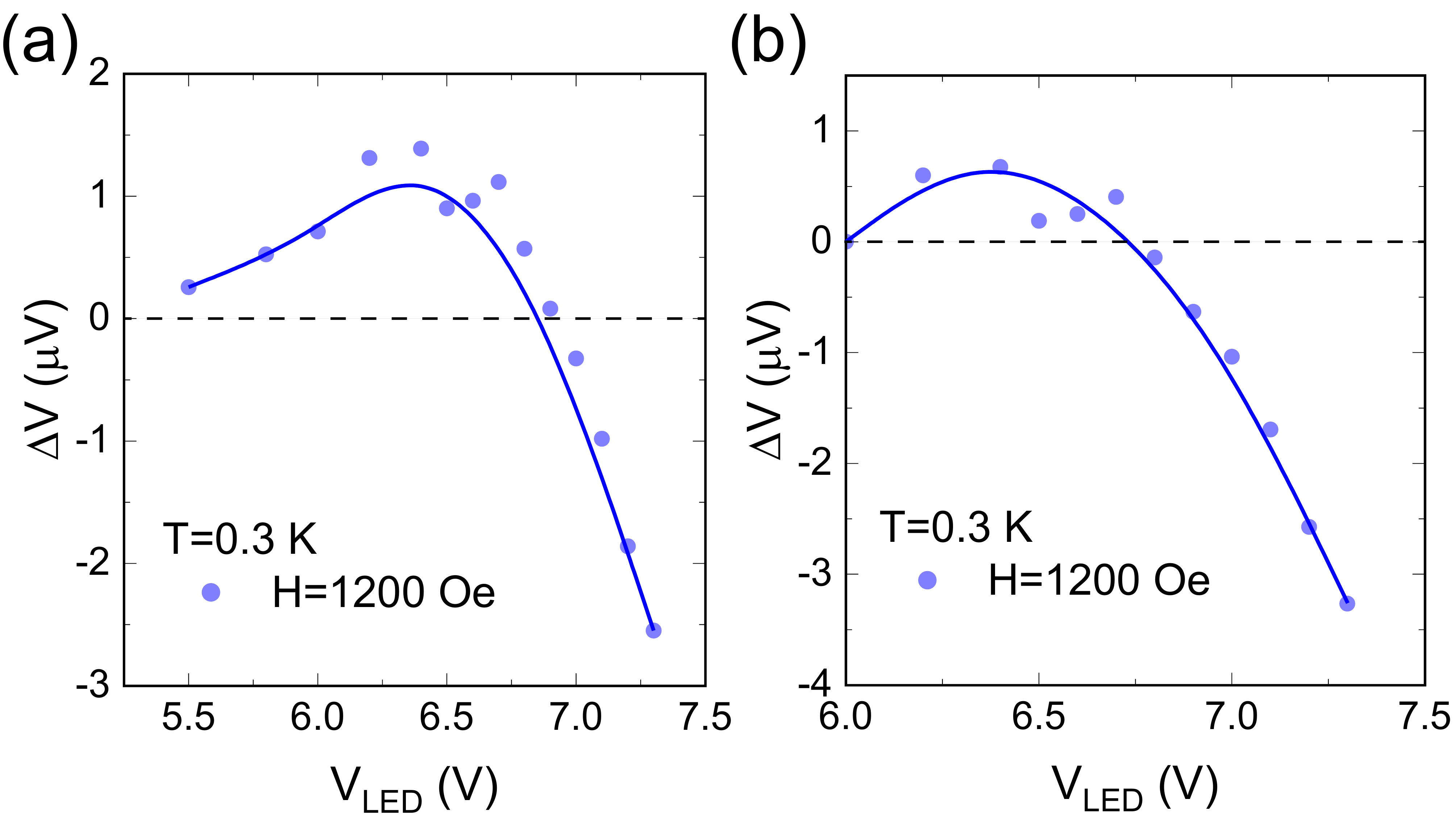}
\caption{(a) TE voltage observed in a S/F/F junction vs LED voltage at $T=0.3$~K ($H=1200$~Oe). 
The voltage background level has not been removed yet. (b) Same TE voltage vs $V_\text{LED}$ curve as in panel (a) after the background voltage subtraction.}
\label{background}
\end{center}
\end{figure}

Some of the plotted TE voltage curves such as figs. \ref{Fig2}(b--c), \ref{Fig3}(a--c) and \ref{Fig4}(a) on the main text were constructed assuming negligible ${\Delta}T$ up until $V_\text{LED}=6$~V, considering that below this voltage practically no temperature gradient was induced between the superconductor and ferromagnet (see Fig. \ref{calibration}(d) inset). Figure \ref{background}(a) shows a typical TE voltage curve against the LED voltage, from 5.5 to 7.3~V. As TE effects seen at voltages below $V_\text{LED}=6$~V are not related with the heat produced by the LED, we assume that they are a consequence of the unavoidable background temperature gradient in the S/F/F junction due to the thermalization process with the cryostat cooling system ($He^3$ pot).
Accordingly, the background voltage was removed on all of the mentioned plots. Fig. \ref{background} displays an example of background TE signal removal for one of the curves.

\section*{4. Modelling of the temperature profile}

The thermal response of the system was modelled using COMSOL \cite{Comsol}. The following section explains the modelling for a fixed voltage of the LED heater of 7~V as a representative example, since this was one of the most used voltage values where the TE response was evaluated.

The LED heater worked at 7~V with an applied current of 0.3~mA. The total dissipated power in this situation is therefore 2.1~mW. The heater is located at a distance of $\sim1$~cm from the sample, so the power per unit area reaching the surface of the sample holder is estimated as 6.7~W/m$^2$. The measured samples surface varies from $30\times30$ to $60\times60$~$\mu\text{m}^2$ depending on the sample, but the structure is grown as a pillar and located under a gold contact with a surface of approx $1\times1$~$\text{mm}^2$. We will first assume that all of the energy that falls onto the pillar is transmitted into the sample's top layer. Therefore the total heat flux entering the sample is taken as $6.7\times10^{-6}$~W. The total temperature gradient had an approximately linear dependence with this total flux and therefore applied voltage of the LED heater.

\begin{figure}
\begin{center}
\includegraphics[width=\linewidth]{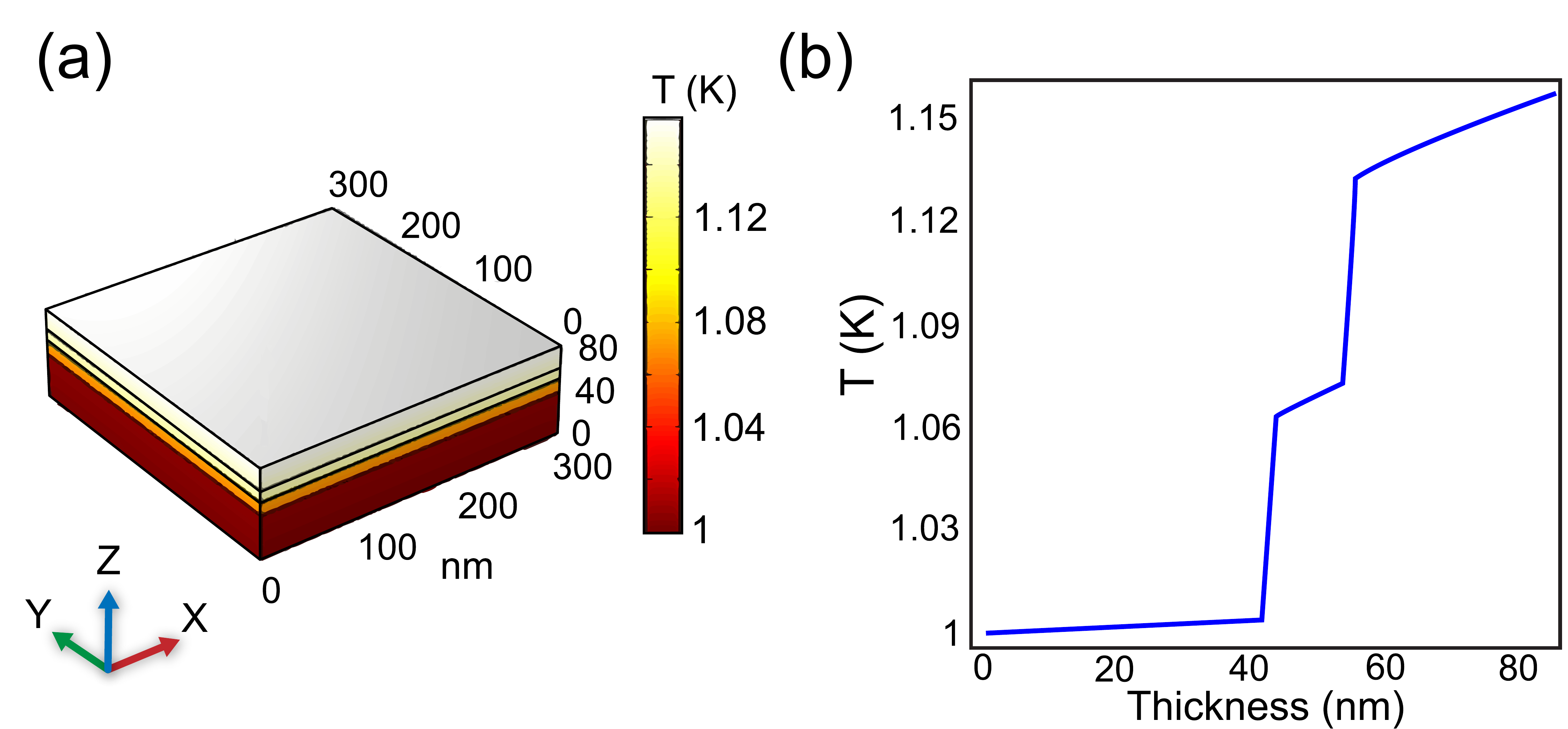}
\caption{COMSOL modelling of the S/F/F system. (a) 3D model of the junction temperature distribution, with a net heat flux of $6.7\times10^{-6}$~W entering from the top. (b) Temperature profile along the Z direction.}
\label{ComsolFig}
\end{center}
\end{figure}

The thermal conductivity and heat capacity of each of the materials (Co, Fe, MgO and V) at low temperatures was taken from tabulated values in the literature \cite{SpecificHeats,MgO,VanadiumK,VanadiumCp}. The model in COMSOL assumes that the bottom of the V layer has a fixed temperature of 1~K and the lateral faces are isolated (since the sample is in a vacuum chamber in the cryostat). With these parameters, the temperature difference between the Fe and V layers is in the range of 70~mK (see Fig. \ref{ComsolFig}). With the observed TE voltage in the order of $\sim3~\mu\text{V}$, this would make for a Seebeck coefficient of $\sim50~\mu\text{V/K}$, which is in the same order of magnitude of previous similar studies \cite{Walter2011}.

\begin{figure}
\begin{center}
\includegraphics[width=\linewidth]{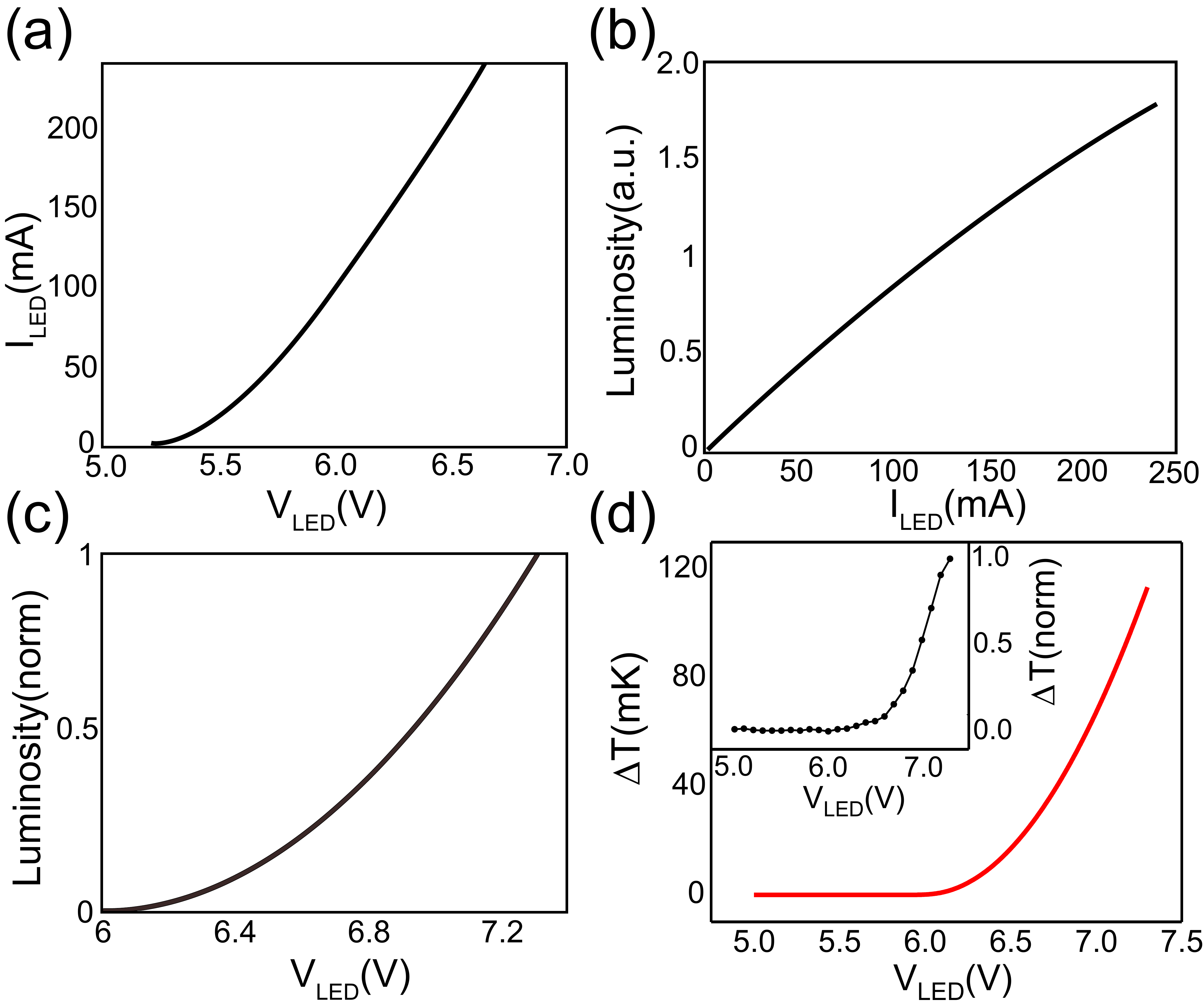}
\caption{(a) Intensity vs applied bias curve at room temperature for the commercial LED that was used during the experiments (model LUXEON 3030 2D). (b) Luminosity vs current curve, for the same LED. (c) Normalized luminosity vs applied voltage curve, constructed by composing the two previous ones and with the voltage range already rescaled to the LED behavior at low temperatures. (d) ${\Delta}T$ vs $V_\text{LED}$ constructed from the curve in (c) plus the COMSOL simulation results. The inset shows the normalized increase in temperature measured experimentally with the thermometer closest to the sample inside the cryostat during a $V_\text{LED}$ sweep. Note that this does not correspond to the temperature gradient ${\Delta}T$ established in the junction, as it is only a local measure near the samples. Nevertheless, the two curves qualitatively match, which is a good indicator for the reasonable validity of the estimated ${\Delta}T$ gradient under the LED heating.}
\label{calibration}
\end{center}
\end{figure}

\section*{5. Rescaling $V_\text{LED}$ into ${\Delta}T$}

Once the temperature gradient was estimated for a fixed bias of $V=7$~V with the COMSOL simulations, the whole ${\Delta}V$ vs. $V_\text{LED}$ curves that were measured for the different studied samples had to be recalibrated into ${\Delta}V$ vs. ${\Delta}T$. For this, first we took a look at the technical datasheet of the LED used in the experiments, from which luminosity vs current (Fig. \ref{calibration}(a)) and current vs. voltage (Fig. \ref{calibration}(b)) curves were obtained. By composing these two curves we obtained a luminosity vs voltage curve our LED. As the curves were obtained at room temperature, the applied voltage axis had to be re-scaled in order to fit the behavior of the LED at low temperatures: for the minimum working voltage, we looked at the temperature values measured by the thermometer during $V_\text{LED}$ sweeps, observing that the LED started heating the sample at an applied bias of $\sim6$~V (vs $5.6$~V at room $T$). As for the maximum luminosity bias value, we observe that the Keithley 228A power supply could not exceed 7.35~V during normal operation when the LED was connected. With this, we can re-scale the luminosity vs $V_\text{LED}$ curve (Fig. \ref{calibration}(c)). The last step is to assume that the heating will be directly proportional (i.e. linear dependence) to the emitted luminosity of the LED, and use 2 points to calibrate the slope: the base point of $V_\text{LED}=6$~V at which ${\Delta}T$ is assumed to be zero, and the simulated point of $V_\text{LED}=7$~V for which an estimation of ${\Delta}T=150$~mK was obtained with COMSOL. Putting all together, we finally have the ${\Delta}T$ vs. $V_\text{LED}$ curve (Fig. \ref{calibration}(d)), which was used to transform the ${\Delta}V$ vs. $V_\text{LED}$ experiments into the presented ${\Delta}V$ vs ${\Delta}T$ graphs.

\section*{6. Thermoelectric response in a short-circuited S/F/F junction}

\begin{figure}[t]
\begin{center}
\includegraphics[width=\linewidth]{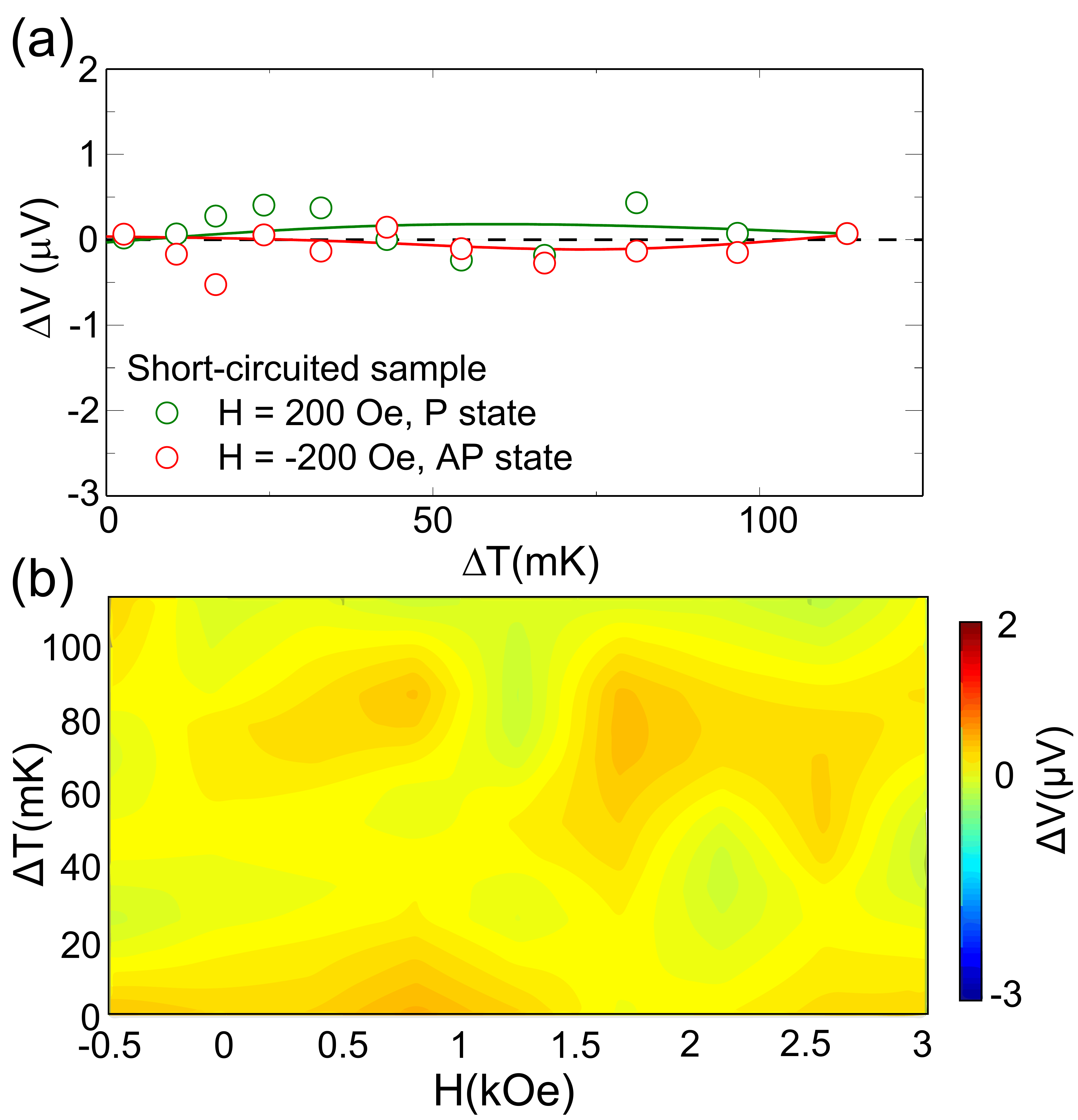}
\caption{Thermoelectric response of a short-circuited S/F/F sample. (a) Thermoelectric voltage for 200~Oe and $-200$~Oe at $T=0.3$~K. (b) 3D representation of the TE voltage against the in-plane field and ${\Delta}T$ at $T=0.3$~K.}
\label{dead_sample}
\end{center}
\end{figure}

We carried out control experiments of the TE response in short-circuited junctions. By operating at relatively high applied biases (between 1 and 2~V), a pinhole was induced in the MgO barriers of one of the samples where the TE response was being studied. The outcome was a close to $10^3$ drop in the resistance, from kOhms to tens of Ohms. The result is shown in Fig. \ref{dead_sample}, revealing an absence of TE generated voltage for any value of the LED heating and magnetic field. The results suggest that the electric-channel was short-circuited, which was accompanied by a strongly enhanced thermal conductivity and therefore a reduced temperature gradient.


\ \\
\ \\

\ali{
    \section*{7. Thermoelectric effect vs. domain-wall state}
    \begin{figure}[b!]
        \centering
        \includegraphics{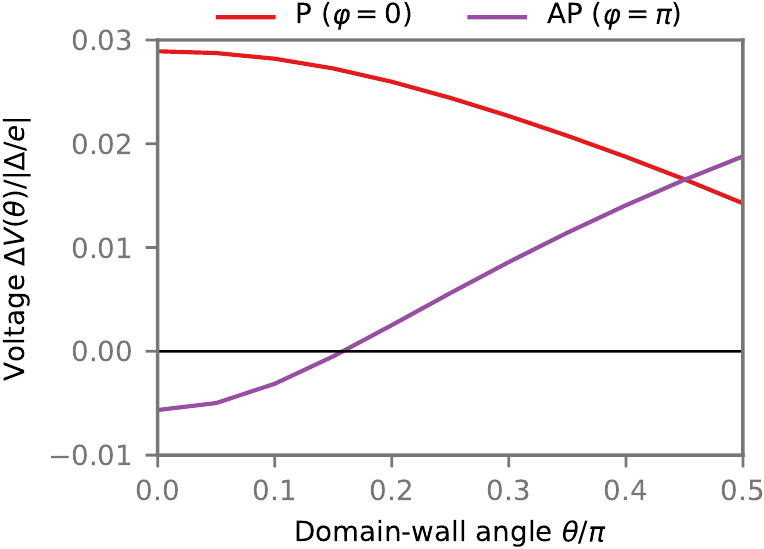}
        \caption{\ali{Thermoelectric response as a function of the interfacial domain-wall angle $\theta$. The other parameters of the junction are the same as in Fig.~\ref{fig:model}(d--e) in the main text.}}
        \label{fig:theta}
    \end{figure}
    As discussed in the main text, we used the Usadel formalism to model the thermoelectric properties of superconducting spin valves. In Fig.~\ref{fig:model}(d--e) of the main text, we showed specifically how the thermoelectric voltage $\Delta V(\varphi)$ behaved as a function of the in-plane misalignment angle $\varphi$ for fixed domain-wall angles $\theta$. Close comparison of the two subfigures shows that in the P configuration ($\varphi = 0$) the thermoelectric voltage has the same sign for both $\theta = 0$ and $\theta = \pi/4$, whereas in the AP configuration ($\varphi = \pi$) the thermoelectric voltage must change sign as a function of $\theta$. In \cref{fig:theta}, we provide a more detailed exposition of this effect, by plotting explicitly $\Delta V(\theta)$ for these two fixed values of $\varphi$. The other junction parameters are the same as in Fig.~\ref{fig:model}(d--e) in the main text. We observe that reorienting the domain wall can flip the sign of the thermoelectric effect for the AP but not the P configuration. Moreover, if the variation in $\theta$ is sufficiently large, then the system can even be tuned between fully symmetric thermoelectric effects ($V_\text{AP} \approx V_{\text{P}}$) and antisymmetric-like thermoelectric effects (opposite signs of $V_\text{AP}$ and $V_\text{P}$).
}

\ali{
    The most direct way to manipulate $\theta$ is to apply external magnetic fields. Applying in-plane fields $H$ should reorient $\theta \rightarrow 0$ such that out-of-plane magnetization components are reduced, and can thus lead to the sign changes in $\Delta V$ that are seen in Fig.~\ref{Fig3} of the main text as $|H|$ is increased. It might also be possible that other system parameters such as junction temperature could affect $\theta$ and thus lead to similar sign reversals in $\Delta V$--especially when this occurs in conjunction with an applied field~$H$. This is one hypothesis for the observed sign changes in Fig.~\ref{Fig3} of the main text as function of $\Delta T$ near the P--AP transitions: When the LED power is increased to produce higher $\Delta T$, the junction's average temperature also increases, and this could affect $\theta$. If $\theta$ is more sensitive to temperature changes for $H$ values bordering a magnetic transition, it might change sufficiently for $\Delta V(\theta)$ to flip sign.
}

\end{document}